%% file: ms.tex
\documentclass{aastex6}

\usepackage{graphicx}

\def\lapp{\ifmmode\stackrel{<}{_{\sim}}\else$\stackrel{<}{_{\sim}}$\fi}
\def\gapp{\ifmmode\stackrel{>}{_{\sim}}\else$\stackrel{>}{_{\sim}}$\fi}
\def\degr{\ifmmode^{\circ}\else$^{\circ}$\fi}

\begin{document}

\title{Pulsar Rotation Measures and Large-scale Magnetic Field Reversals in the Galactic Disk}
\author{J. L. Han\altaffilmark{1}}
\author{R. N. Manchester\altaffilmark{2}}
\author{W. van Straten\altaffilmark{3}}
\author{P. Demorest\altaffilmark{4}}
\altaffiltext{1}{National Astronomical Observatories, Chinese Academy of 
	Sciences, Jia 20 DaTun Road, Beijing 100012, China. 
	Email: hjl@bao.ac.cn}
\altaffiltext{2}{CSIRO Astronomy and Space Science, PO Box 76,
        Epping, NSW 1710, Australia. Email: dick.manchester@csiro.au}
\altaffiltext{3}{Institute for Radio Astronomy and Space Research, Auckland University of Technology, PB 92006, Auckland 1142, New Zealand}
\altaffiltext{4}{National Radio Astronomy Observatory, PO Box O, Socorro, NM 87801, USA}

\begin{abstract}
We present the measurements of Faraday rotation for 477
pulsars observed by the Parkes 64-m radio telescope and the Green Bank
100-m radio telescope.  Using these results along with previous
measurements for pulsars and extra-galactic sources, we analyse the
structure of the large-scale magnetic field in the Galactic
disk. Comparison of rotation measures of pulsars in the disk at
different distances as well as with rotation measures of background radio
sources beyond the disk reveals large-scale reversals of the field
directions between spiral arms and interarm regions. We develop a
model for the disk magnetic field, which can reproduce not only these
reversals but also the distribution of observed rotation measures of
background sources.
\end{abstract}

\section{Introduction}\label{sec:intro}
Interstellar magnetic fields of our Galaxy have long been known to
play fundamental roles in astrophysics and astroparticle physics, and
their properties have been investigated for many years. A 
Galactic magnetic field was proposed by \citet{fer49} as the agent for
transport of cosmic rays through interstellar space and, shortly
afterward, \citet{kie50} proposed a synchrotron origin for the
Galactic background of radio emission. Remarkably, both Fermi and
Kiepenheuer calculated the strength of the interstellar field to be of
order a few $\mu$G, very close to current estimates. Magnetic fields
contribute significantly to the interstellar hydrodynamic pressure
\citep{bc90} and may even be dynamically important in the outer parts
of some galaxies \citep{bf07}. The strong magnetic fields found in
molecular clouds are key to understanding the star-formation process
\citep{ree87}. Understanding the structure of the Galactic magnetic
field is also important to understanding the origin and maintenance of
magnetic fields in other galaxies and in intergalactic space
\citep{bbm+96}. For a recent review of Galactic and extragalactic
magnetic field observations see \citet{han17}. 

Several tracers have been used to investigate interstellar magnetic
fields, including starlight polarization
\citep[e.g.,][]{hei96,cppt12}, Zeeman splitting of spectral lines of
HI and various molecules \citep[e.g.,][]{cru99,vle08}, background
synchrotron radiation from our Galaxy
\citep[e.g.,][]{bkb85,blw+13,paa+16a}, polarized thermal emission from
dust grains in molecular clouds \citep[e.g.,][]{ncr+03,paa+16b} and
Faraday rotation of extragalactic radio sources (EGRS)
\citep{sk80,tss09} and of pulsars \citep{man72,rl94,hmq99,
  hml+06,njkk08}. These and similar observations have shown that the
large-scale magnetic field in galactic disks is largely toroidal and
aligned with spiral arm structures, whereas halo fields probably have
azimuthal fields with reversed directions above and below the Galactic
plane \citep{hmbb97,hmq99} though the field scale-height and
scale-radius are not yet known. Polarisation observations of
synchrotron emission from nearby galaxies suggest that large-scale
magnetic fields in galactic disks are predominantly spiral with
roughly the same pitch angle both within spiral arms and in interarm
regions \citep[e.g.][]{bec15}. 

Pulsars are very effective probes of the magnetic field of our Galaxy
\citep{man74,ls89,rk89,wck+04,hml+06}. They are highly polarized, have
no intrinsic Faraday rotation and are widely distributed throughout
the Galaxy at approximately known distances, allowing a
three-dimensional tomographic analysis of the field
structure. Furthermore, the pulse dispersion gives a unique 
calibration of the integrated electron density along the line of
sight, allowing a direct estimate of the strength of the field:
\begin{equation}\label{eq:bpar}
  \langle B_{||}\rangle = \frac{\int_0^D n_e {\bf B}.d{\bf l}}{\int_0^D n_e dl}
  = 1.232\frac{\rm RM}{\rm DM},
\end{equation}
where $\langle B_{||}\rangle$ is the mean line-of-sight magnetic field component in $\mu$G,
weighted by the local electron density $n_e$, $D$ is the pulsar distance
and RM and DM are respectively the pulsar rotation measure and
dispersion measure in their usual units (rad~m$^{-2}$ and cm$^{-3}$~pc). By
using pairs of pulsars that are close together on the sky and at
distances $D_1$ and $D_2$ respectively, the mean line-of-sight field
component between $D_1$ and $D_2$ can be obtained from:
\begin{equation}\label{eq:dbpar}
  \langle B_{||}\rangle_{D_2 - D_1} = 1.232\frac{\rm RM_2 - RM_1}{\rm DM_2 - DM_1}.
\end{equation}
This relation allows analysis of changes in the magnetic field along a
given line of sight and hence full tomographic mapping of the Galactic
magnetic field. \citet{bssw03} argued that correlated or
anti-correlated electron density with field strength will
strongly bias estimates of $B_{||}$, but detailed simulations by
\citet{wkr+09,wkr15} show this not to be a problem, although the
uncertainty of estimated mean field strengths depends on the Mach
number of the interstellar medium.

In addition to the field strength, the field directions and their
reversals are crucial to the understanding of the structure of
Galactic large-scale magnetic fields.  Many authors have proposed
models for the large-scale structure of the Galactic magnetic
field. Based on the RMs of 38 relatively nearby pulsars, \citet{man74}
concluded that the local Galactic magnetic field is basically
azimuthal and directed toward longitude $l\sim 90\degr$, that is,
clockwise when viewed from the Galactic North Pole. \citet{tn80}
analyzed RMs of 48 pulsars and found evidence for a field reversal in
the inner Carina-Sagittarius arm.  After a large number of pulsar RMs
were obtained by \citet{hl87}, \citet{ls89} confirmed the first inner
reversal and suggested another reversal in the outer
Galaxy. \citet{rk89} and \citet{rl94} fitted concentric ring models to
the pulsar RM data with alternating field directions in each
ring. Based on more than 350 pulsar RMs, \citet{val05} revised the
ring model to have a dominant clockwise ring field with just one ring
of counter-clockwise field in the Galactocentric radius range 5 --
7~kpc. Earlier, both an axisymmetric spiral (ASS) model \citep{val96}
and a bisymmetric spiral (BSS) model in which alternate arms have
oppositely directed fields \citep{hq94,id99,hmq99} for the global disk
magnetic field were suggested, but more recent pulsar RM data do not
favour this interpretation \citep{mfh08}. \citet{njkk08} attempted to
fit more complex bisymmetric three-dimensional models to pulsar RM
data, but the results were inconclusive with none of the tested models
giving a good fit to the data.

Radio continuum surveys have enabled measurement of RMs for thousands
of EGRS \citep{tss09}, and many authors have used these RMs to
constrain models for the large-scale structure of the Galactic
magnetic field. Some modelling \citep{sk80,sf83,ptkn11} favoured a
bisymmetric model for the disk field but with only one or two
identified field reversals. Most recent models
\citep{bhg+07,srwe08,sr10,vbs+11,jf12} with the benefit of a larger
sample of EGRS RMs \citep{btj03,vbs+11} are dominated by a general
clockwise field but with counterclockwise fields in the
Sagittarius/Scutum -- Crux spiral zone. Recently, \citet{obkl17} used
both EGRS and Galactic continuum background data to suggest that the
outer boundary of this spiral zone is not perpendicular to the
Galactic plane but is sloping toward later longitudes at positive
latitudes and earlier longitudes at negative latitudes.

RMs for EGRS are of course integrated along the entire ray path
through the Galaxy. They are therefore less sensitive to reversals in
the Galactic field direction than pulsars which are distributed
throughout the Galaxy. Also EGRS have an intrinsic RM component from
Faraday rotation in the host galaxy and also a possible intergalactic
component. In analyses of EGRS RMs, these components are generally
assumed to be random, just adding to the fluctuations from small-scale
variations in the Galactic magnetic field.

Using pulsar RMs, \citet{hml+06} concluded that the data were best
represented by a model in which spiral arms have a counter-clockwise
field and interarm regions have a clockwise field. This is similar to
the BSS models, but with twice as many reversals. This idea has
received support from \citet{nk10} who found evidence from both pulsar
and EGRS RMs for a clockwise interarm field between counter-clockwise
fields in the Norma and Crux spiral arms.

In the last decade the NE2001 Galactic electron density model
\citep{cl02} has been widely adopted to estimate pulsar distances and
also used in modelling of the Galactic magnetic field from EGRS RMs.
In this paper, we use the new YMW16 Galactic electron density model
\citep{ymw17} to estimate pulsar distances from DMs because this model
is believed to give more reliable estimates in general \citep[see
  Table 6 of][]{ymw17}. Even so, estimated distances to some pulsars
can be in error by a factor of two or even more.

Currently there are 732 published pulsar RMs
\citep[see][]{mhth05}\footnote{\url{
 http://www.atnf.csiro.au/research/pulsar/psrcat}, V1.56} of which two
are for pulsars in the Small Magellanic Clouds, so we have 730
previously published Galactic RMs. There are nearly 2600 Galactic
pulsars in the ATNF Catalogue, so there is much scope for new pulsar
RM determinations. In this paper, we present measurements made using
the Parkes 64-m radio telescope and the Robert C. Byrd Green Bank
Telescope (GBT) in several sessions in 2006 and 2007. About 500
pulsars were observed at Parkes in the 20cm band ($\sim$1400 MHz) and
about 125 pulsars were observed using the GBT in the 35cm band
($\sim$800 MHz). Analysis of these observations resulted in the
determination of RMs for 477 pulsars, of which 441 are either new or
more precise than previous measurements.
We combine these new measurements with previously published pulsar RMs
and with RMs of EGRS to investigate the large-scale structure of the
magnetic field in the Galactic disk. Our observations and data
reduction methods and the RM samples that we use are described in
\S\ref{sec:obs}. Details of the analysis for large-scale magnetic
fields in different zones of the Galactic disk are given in
\S\ref{sec:tgtB}. In \S\ref{sec:model} we describe a simple model for
the Galactic disk magnetic field that is consistent with the Galactic
field structures including arm/interarm reversals that we find and
with the distribution of EGRS RMs. We conclude the paper in
\S\ref{sec:concl} with a brief summary of the main results and the
prospects for future work.

\section{Observations, data processing methods and the RM samples}
\label{sec:obs}

\subsection{Pulsar observations}
We used the Parkes 64-m and the Green Bank 100-m telescopes to observe
pulsars that had no previously measured rotation measure (RM) but were
sufficiently strong to have a reasonable prospect of measuring a
significant RM in one hour or less for Parkes or 15 minutes or less for
the Green Bank Telescope (GBT). A few strong pulsars with well-known
RMs were also observed at the start of each session as system
checks. 

The Parkes observations were made in seven sessions between 2006
August 2006 and 2008 February. All observations were in the 20-cm band
and, except for one session (2007 March), all used the central beam of
the 13-beam multibeam receiver \citep{swb+96} with a central frequency
of 1369 MHz and an observed bandwidth of 256 MHz. For the 2007 March
observations the ``H-OH'' receiver was used with the same bandwidth
but at a central frequency of 1433 MHz. Both systems receive
orthogonal linear polarisations and have a pulsed calibration signal
injected at $45\degr$ to the two feed probes. The system-equivalent
flux densities for the two receivers were about 35 Jy and 42 Jy
respectively, determined using calibration observations on and off the
strong radio source Hydra A, assumed to have a flux density of 43 Jy
at 1400 MHz and a spectral index of $-0.91$ \citep{bgpw77}. Multibeam
observations were made with half the total observing time at each of
two feed angles, $\pm 45\degr$, to reduce the effect of feed
cross-coupling on the results. This was not necessary for the H-OH
receiver. In the 2006 and 2007 sessions, data were recorded using the
PDFB1 signal-processing system; for the two 2008 sessions the PDFB2
system was used. Both systems used a polyphase filterbank and produced
mean pulse profiles in 1-minute sub-integrations with full
polarisation data in each of 512 frequency channels and with either
512 or 1024 bins across the pulse period. A brief description of these
systems is given by \citet{mhb+13}. Data were stored for subsequent
analysis as PSRFITS files \citep{hvm04}.

The GBT observations were made in 2007 November using the 800~MHz
prime focus receiver \citep[see][for details]{hdvl09} which has a
system-equivalent flux density on cold sky of approximately 15~Jy. The
Green Bank Astronomy Signal Processor (GASP) pulsar observing system
\citep{dem07,fsb+04} was used with a central frequency of 774 MHz and
a bandwidth of 96 MHz. Flux density calibration was via 3C286 and
3C295, with assumed flux densities at 774 MHz of 19.44~Jy and 35.45~Jy
respectively. A polyphase filterbank was used to divide the signal
into 4-MHz sub-bands which were distributed to a 16-node computer
cluster for real-time coherent dedispersion and additional frequency
division to a final resolution of 0.25~MHz. Dedispersed data in each
sub-band were then folded in real-time into 1024 pulse phase bins for
30-s sub-integrations and stored using the PSRFITS data format. A
1-min pulsed calibration observation and two 4-min observations at
orthogonal feed angles were made for each pulsar.

\subsection{Analysis methods for rotation measures}

Off-line data analysis including polarimetric calibration and RM
determination \citep[see section 2 of][for details for GBT
  observations]{hdvl09} was performed using the {\sc psrchive} pulsar data
processing system \citep{hvm04}. First, the frequency-time data were
examined for radio frequency interference and affected data were
excised. Next the data were calibrated to compensate for instrumental
gain and phase variations across the band, converted to Stokes
parameters and placed on a flux density scale. Where applicable, the
two observations at orthogonal feed angles were then summed and the
whole observation summed in time. From the resulting multi-frequency
polarisation profiles, the RM of each pulsar was obtained as follows.
A first guess at the RM was found by searching the range of $\pm 2000$
rad~m$^{-2}$ for a peak in the total linear polarisation $L = (Q^2 +
U^2)^{1/2}$ summed across all on-pulse phase bins and across the band,
where $Q$ and $U$ are the linear Stokes parameters. This value was
then iteratively refined by taking the current best estimate of the RM
and summing the data separately in the two halves of the band. A
correction to the RM was then obtained from a weighted mean
position-angle difference across the pulse profile. Taking the
weighted mean difference makes the process relatively immune to
orthogonal mode transitions \citep[cf.][]{rbr+04}. The final RM value
was then obtained by subtracting the ionospheric RM contribution to
give the RM along the path from the top of the ionosphere to the
pulsar. The ionospheric RM was computed using a model for the
geomagnetic field and the International Reference Ionosphere 2007
\citep{br08}. It was typically between 0.4 to 3.5 rad~m$^{-2}$ for GBT
observations and $-0.2$ to $-2.0$ rad~m$^{-2}$ for Parkes
observations, with a largely diurnal variation.

\input tab1.tex

\input tab2.tex

\subsection{Rotation measure samples}\label{sec:rmobs}

Table~\ref{tb:psrrm} lists 501 RM measurements for 477 pulsars; for 21
pulsars, repeated observations were made in different sessions, either
as a system check or in an attempt to get a better RM measurement.
Columns 1 -- 6 list the pulsar J2000 name, period, dispersion measure,
Galactic longitude, Galactic latitude and the estimated pulsar
distance. Unless independent distance estimates are available,
distances are based on the YMW16 electron-density model \citep{ymw17}.
The next two columns list the measured RM and its uncertainty, and the
final two columns give the telescope and date of the observations.
For the pulsars with repeated measurements, we formed weighted mean
RMs for use in subsequent analysis. These are listed in
Table~\ref{tb:meanrm}.

The ATNF Pulsar Catalogue (V1.56) lists published RMs for 732 pulsars,
of which two are for pulsars that lie in the Small Magellanic Cloud.  A
total of 91 pulsars in Table~\ref{tb:psrrm} have previously published
RM measurements and so the total number of available Galactic RMs is
1116. We compare RM values with the previously published values in
Table~\ref{tb:pubrm}. The first two columns give the pulsar J2000 name
and the B1950 name if one exists and the next two columns give our
best RM measurement from Table~\ref{tb:psrrm} or
Table~\ref{tb:meanrm}. The following columns are grouped in sets of
three, with the first two columns giving a previously published RM and
its quoted uncertainty and the third column giving the reference key
for the publication. More recent publications are listed first and
reference keys are identified in the Table footnote. Where a reference
key for an earlier paper is marked with an asterisk, the corresponding
RM measurement is evidently the best available. This applies to 36 of
the 91 pulsars and these values are used in subsequent analyses.

\input tab3.tex

Most of the new measurements are in good agreement with previously
published values with only ten cases where the RM difference exceeds
five times the combined uncertainty. Some of the smaller differences
are likely to result from temporal variations of the RM as seen in,
for example, the Vela pulsar \citep{hhc85}. Four of the measurements
are discrepant by more than $10\sigma$. The largest of these
discrepencies is for PSR J1707$-$4053 where our measured RM is
$-183.7\pm 3.4$~rad~m$^{-2}$ compared to $+168.0\pm 4.0$~rad~m$^{-2}$
from \citet{njkk08}. Our measurement agrees well with a more recent
measurement by \citet{fdr15} and with an older measurement by
\citet{qmlg95} and so it appears that the \citet{njkk08} measurement
is incorrect. The other three cases are PSRs J1623$-$4256,
J1836$-$1008 and J1935+1616.  Previous observations for PSR
J1623$-$4256 \citep{hml+06} unfortunately were processed with an incorrect DM;
re-analysis of these data with the correct DM gives an RM
consistent with that presented here. For PSR J1836$-$1008, the new RM
in Table~\ref{tb:psrrm} is confirmed by analysis of more recent data.
For PSR J1935+1616, the new measurement agrees with two previous
determinations, so it appears that the \citet{jhv+05} result is
discrepant.  Among the 1116 Galactic pulsars with RMs, 787 are at low
Galactic latitudes ($|b|<8\degr$) and hence most relevant to the
present work.

We also make use of 3933 RMs of extragalactic radio sources (EGRS)
with Galactic latitude $|b| < 8\degr$
\citep{xh14}.\footnote{\url{http://zmtt.bao.ac.cn/RM/}} Of these, 2942
are from the analysis of NVSS data by \citet{tss09}, 283 from
\citet{btj03}, 184 from \citet{vbs+11}, and 104 from
\citet{bhg+07}. The remaining 298 RMs are from a variety of papers.

\section{Large-scale field structure in the Galactic disk}\label{sec:tgtB}

\begin{figure*}
\centering\includegraphics[bb = -45 -45 402 402, clip, width=0.95\textwidth]{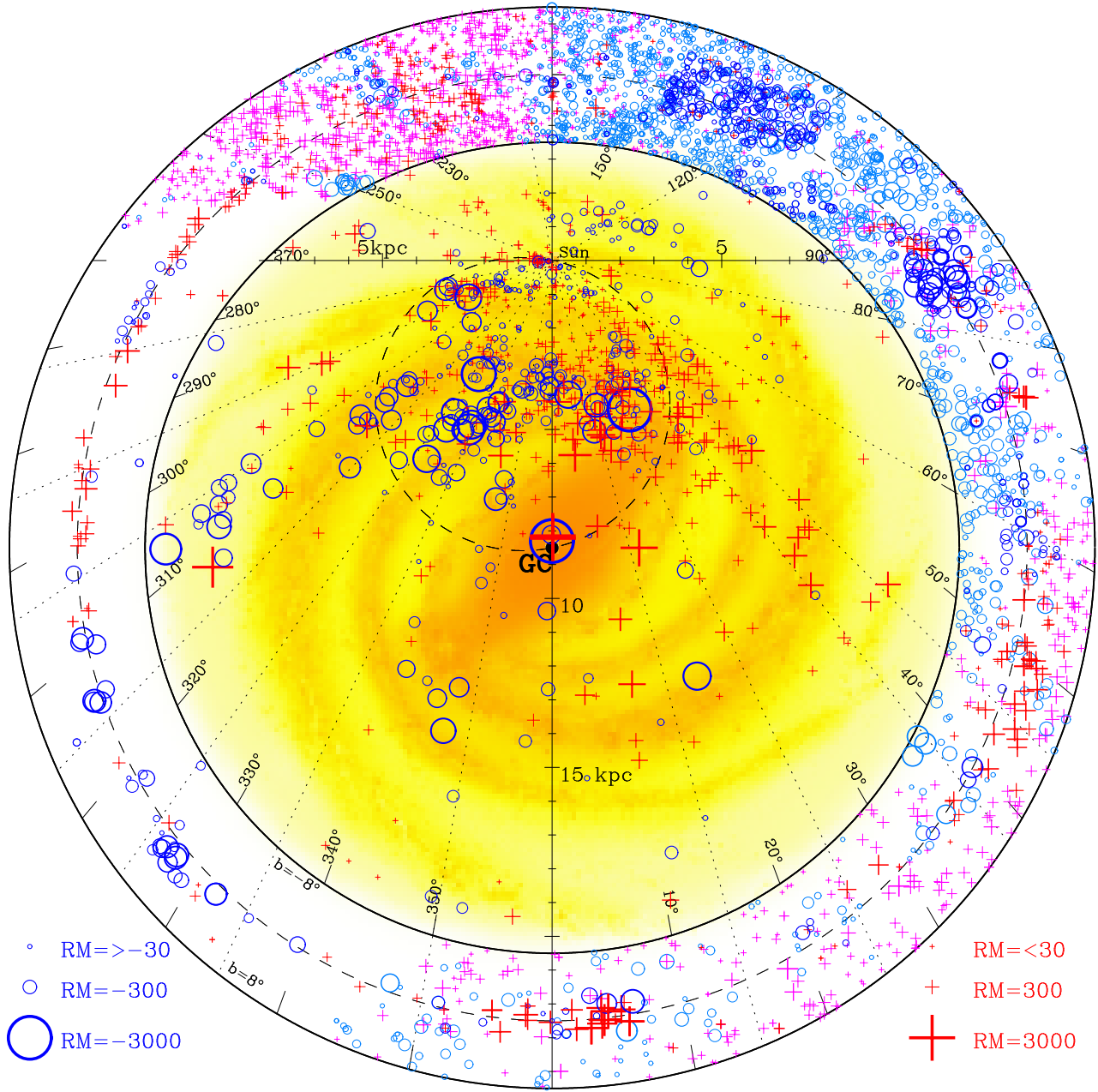}
\caption{Distribution of pulsars and EGRS with measured RMs in
  Galactic coordinates. The central part of the figure shows RMs for
  pulsars with Galactic latitude $|b|<8\degr$ projected on to the
  Galactic disk and the outer ring shows the distribution of RMs for
  EGRS with $|b|<8\degr$. Radial distance in the outer ring is
  linearly proportional to Galactic latitude. The light blue and pink
  symbols indicate the RMs derived from NVSS data \citep{tss09}. Blue
  circles and red + signs in both central part and the outer ring
  indicate negative and positive RMs respectively and the symbol size
  is proportional to $|RM|^{1/2}$. The background in the central part
  of the figure is an artist's impression of the Galactic structure
  modified from a NASA/JPL image (Credit: R. Hurt) according to
  updated spiral arms of \citet{hh14}. The dashed circle gives the
  locus of tangent points assuming a spiral pitch angle of $11\degr$
  and the dotted lines give the longitude boundaries of the
  approximately tangential spiral-arm or interarm zones and
  anti-center zones used for the RM analyses in this
  paper.}\label{fg:rm_disk}
\end{figure*}

Figure~\ref{fg:rm_disk} shows the overall RM distribution for both
pulsars and EGRS within $8\degr$ of the Galactic plane. In general
terms, there is predominance of positive RMs in the first and third
Galactic quadrants (i.e., $0\degr < l < 90\degr$ and $180\degr < l <
270\degr$) and of negative RMs in the other two quadrants. Since
positive RMs indicate fields directed toward us, overall these results
suggest clockwise fields in the outer Galaxy and counterclockwise
fields inside the Sun viewed from the north Galactic pole.  At least
in the first and fourth quadrants, there is a tendency for more
distant pulsars to have larger RMs, indicating the large scale of the
counter-clockwise fields in the inner Galaxy.

Closer examination of Figure~\ref{fg:rm_disk} shows however that this
counter-clockwise field in the inner Galaxy is predominantly confined
to the spiral arms. This is most clearly revealed by increasingly
negative RMs in the vicinity of spiral-arm tangential points, for
example, the Crux tangential region near Galactic longitude $l \sim
310\degr$ and the Norma tangential region near Galactic longitude $l
\sim 330\degr$.  In contrast, in the interarm tangential regions, for
example, in the Norma-Crux interarm region near $l \sim 320\degr$ and
the Crux-Carina interarm region ($l \sim 300\degr$), RMs are positive,
indicating a clockwise field similar to the local interarm
region. Furthermore, EGRS RMs in the direction around the Crux-Carina
interarm region ($l \sim 300\degr$) are nearly all large and positive,
confirming the large-scale clockwise field. In the first quadrant, the
counter-clockwise fields in the spiral arms are generally clear, but
it is difficult to identify the direction of the interarm fields since
the spiral arms are much closer and less well defined in this
quadrant.

In the following sub-sections, we quantify these RM trends by fitting
the observed variations of RM with DM and distance over specified
distance ranges and directions and comparing these fits with the mean
EGRS RM in the same direction.  As discussed in \S\ref{sec:intro}, we
can use Equation~\ref{eq:dbpar} to give the mean line-of-sight
component of the interstellar magnetic field, weighted by the local
$n_e$, over different distance intervals along the line of sight to
pulsars in similar directions. Distances to individual pulsars derived
from Galactic $n_e$ models are subject to unpredictable
errors. Therefore, rather than fitting to pulsar pairs individually,
we fit linear trend lines to plots of RM vs distance over specified
distance intervals and to plots of RM vs DM over DM intervals that
match the distance range as closely as possible. The averaging over
groups of pulsars minimises the effects of small-scale B-field
fluctuations and distance errors. We emphasize that the derived
B-field estimates are derived solely from the RM -- DM fits.

In order to improve the reliability of the $B$-field estimates, we
omitted RMs of 15 pulsars with an uncertainty larger than
35~rad~m$^{-2}$ and used the Maximum Likelihood Robust Estimate
routine from \citet[][see pp.694-700]{ptvf96} to fit a line by
minimizing absolute deviation (i.e. the {\sc medfit} subroutine). This
``robust'' fitting is necessary so that outliers resulting from HII
regions, other unmodelled electron-density fluctuations or
magnetic-field fluctuations along the path do not unduly influence the
slope of the fitted line. We take as the uncertainty of the slope the
mean absolute deviation of RMs from the fitted line divided by the DM
range for the fitting. Generally, the scatter around the fitted lines
in the RM -- DM plots is dominated by real fluctuations in the
line-of-sight magnetic field components, not RM measurement errors
which are very small compared to the data scatter. Positive slopes of RM vs
DM correspond to magnetic fields directed toward us, i.e., to
clockwise fields in the Galactic Quadrant 4 and counter-clockwise
fields in the Galactic Quadrant 1.

The regions for which we have analysed the RMs are listed in
Table~\ref{tb:tgt}. The arm and interarm designations are guided by
the 4-arm spiral model of \citet[][i.e. the background images of
  Fig.~\ref{fg:rm_disk}]{hh14}, and the ranges for the Galactic
longitude $l$ are chosen accordingly. For the inner Galactic
quadrants, the fitted regions are guided by the tangential zones since
spiral fields have a small angle to our line of sight and distance
errors have less effect there. However, where clear trends in RM
versus DM or distance exist, the fitted region has been adjusted to
encompass these. For the outer Galactic regions, the fitted regions
are determined by the RM trends. The fifth column of
Table~\ref{tb:tgt} gives the number of pulsars in the fitted region or
the number of EGRS for each longitude range. The next two columns give
estimates of $B_{||}$ and field direction from the RM -- DM fits in
the vicinity of tangential regions of Quadrants 1 and 4.  Comparison
of the RMs of background EGRS with the RMs of the most distant pulsars
in each zone gives a good indication of the magnetic field orientation
beyond the pulsars, but it is not possible to reliably estimate field
strengths in these cases because of the uncertain DM contribution.

\begin{deluxetable}{lcllccrDD}
\tabletypesize{\footnotesize}
\tablecaption{Galactic disk zones and their magnetic fields\label{tb:tgt}}
\tablehead{
\colhead{Region } & \colhead{$l$--Range} & \colhead{D--Range} & \colhead{DM--Range} & No. PSRs
 &\colhead{$B_{||}$} & B-field & \multicolumn2c{Arrow $l$} & \multicolumn2c{Arrow D}  \\
                 & \colhead{($\degr$)}   & \colhead{(kpc)}       & \colhead{(cm$^{-3}$~pc)} & or EGRS
 &\colhead{($\mu$G)} & Direction &  \multicolumn2c{($\degr$)}  & \multicolumn2c{(kpc)}      
}
\decimals
\startdata
\multicolumn{9}{c}{Quadrant 1}\\ \hline
Near 3-kpc     & 15 -- 25  & 3.5 --  6.5 & 350 -- 850 & 25 &$+4.0\pm0.7$& ccw &  20 & 5.5 \\
Near 3-kpc -- EGRS & 15 -- 25 & 6.5 -- E  & 850 -- E  & 71 &            &  cw &  20 & 11.0 \\[1mm]
Scutum         & 25 -- 38  & 4.0 -- 8.0 & 200 -- 800  & 46 &$+0.4\pm0.4$& ccw  & 32  & 7.0 \\
Scutum -- EGRS  & 25 -- 38  & 9.5 -- E   & 900 -- E   & 78 &            & -- & -- & -- \\[1mm]
Scutum -- Sgr  & 38 -- 45  & 4.0 -- 12.0 & 200 -- 500 & 25 &$+3.3\pm0.9$& ccw &  42 & 8.5 \\
Scutum-Sgr -- EGRS & 38 -- 45& 12.0 -- E   & 500 -- E & 37 &            &  cw &  42 & 13.0 \\[1mm]
Sagittarius    & 45 -- 60  & 3.0 -- 8.5 & 100 -- 300  & 30 &$+1.4\pm1.0$& ccw &  50 & 5.0 \\
Sagittarius -- EGRS & 45 -- 60& 8.5 -- E   & 300 -- E & 176&            &  cw &  50 & 8.5 \\[1mm]
Local -- Perseus & 60 -- 80  & 3.5 -- 8.0 &  70 -- 250& 14 &$+0.7\pm0.7$& ccw &  73 & 7.0 \\
Local--Perseus -- EGRS & 60 -- 80& 8.0 -- E   & 250 -- E & 225 &            &  cw &  73 & 10.5 \\ 
\hline 
\multicolumn{9}{c}{Outer Zones for the local region and the Perseus arm}\\ \hline
Local Q1-Q2    & 80 -- 120 & 1.0 -- 5.0 &  10 -- 200  &18 &$-1.4\pm0.6$&  cw & 105 & 2.5 \\
Local Q1-Q2 -- EGRS & 80 -- 120& 5.0 -- E  &  200 -- E&576  &           & ccw & 105 & 4.0 \\[1mm]
Outer Q2       & 120 -- 190 &   --   &    --      &   &          & --    &  --   & --   \\
Outer Q3       & 190 -- 250& 0.0 -- 3.5 &   0 -- 130  &13 &$+1.3\pm0.4$&  cw & 235 & 2  \\
Outer Q3 -- EGRS & 190 -- 250& 3.5 -- E   & 130 -- E  &841 &            & ccw & 230 & 3.5 \\[1mm]
Local Q3-Q4    & 250 -- 270& 0.0 -- 6.0 &  30 -- 280 & 20 &$+1.1\pm0.5$&   cw & 260 & 4.0 \\
Local Q3-Q4 -- EGRS & 250 -- 270 &6.0 -- E   & 280 -- E &138 &         &   ccw & 260 & 6.5 \\[1mm]
Outer Carina   & 270 -- 282 &  0.1 -- 1.1  & 50 -- 250 & 23 & $+0.8\pm0.5$ &  cw  & 276  & 0.7 \\ 
Outer Carina -- EGRS & 270 -- 282&  --     & 250 -- E  & 26 &             & ccw & 276 & 9.0 \\ 
\hline
\multicolumn{9}{c}{Quadrant 4}\\ \hline
Carina       & 282 -- 294& 2.0 -- 4.0 & 250 -- 550 &22 &$-1.2\pm1.0$& ccw & 288 & 3.0 \\
Carina -- EGRS  & 282 -- 294& 4.0 -- E  & 550 -- E & 8 &            &  cw &  288 & 11.0 \\[1mm]
Carina -- Crux & 294 -- 304& 2.0 -- 10.0 & 100 -- 600 & 21 &$+1.9\pm0.3$&  cw & 299 & 7.0 \\
Carina--Crux -- EGRS& 294 -- 304& 10.0 -- E & 600 -- E &13 &        &   ccw & 299 & 12.0 \\[1mm]
Crux           & 304 -- 316& 4.0 -- 13.0& 200 -- 800 & 38 &$-1.6\pm0.5$& ccw & 310 & 7.5 \\
Crux -- EGRS    & 304 -- 316& 13.0 -- E  & 800 -- E  &13 &            &  cw & 310 &13.0 \\[1mm]
Crux -- Norma  & 316 -- 325& 3.0 -- 11.0 & 250 -- 700 & 9 &$+1.3\pm0.3$&  cw & 320 &6.0 \\
Crux-Norma -- EGRS&316 -- 325&11.0 -- E   & 700 -- E  & 6 &            & ccw & 320 &12.0 \\[1mm]
Norma          & 325 -- 335& 4.0 -- 6.5 & 300 -- 800 & 15 &$-3.7\pm0.6$& ccw & 330 & 6.5 \\
Norma -- EGRS   & 325 -- 335&10.0 -- E   & 900 -- E  & 20 &            &  cw & 330 &12.5 \\[1mm]
Far 3-kpc      & 335 -- 350& 8.0 -- 13.5 & 400 -- 700 & 23 & $-3.1\pm1.1$ & ccw & 343 &12.5 \\
Far 3-kpc -- EGRS   & 335 -- 350&13.5 -- E & 700 -- E & 23 &              &  cw & 343 &15.5
\enddata
\end{deluxetable}

\subsection{Fourth Galactic Quadrant}\label{sec:q4}

\begin{figure}
\includegraphics[angle=270,width=130mm]{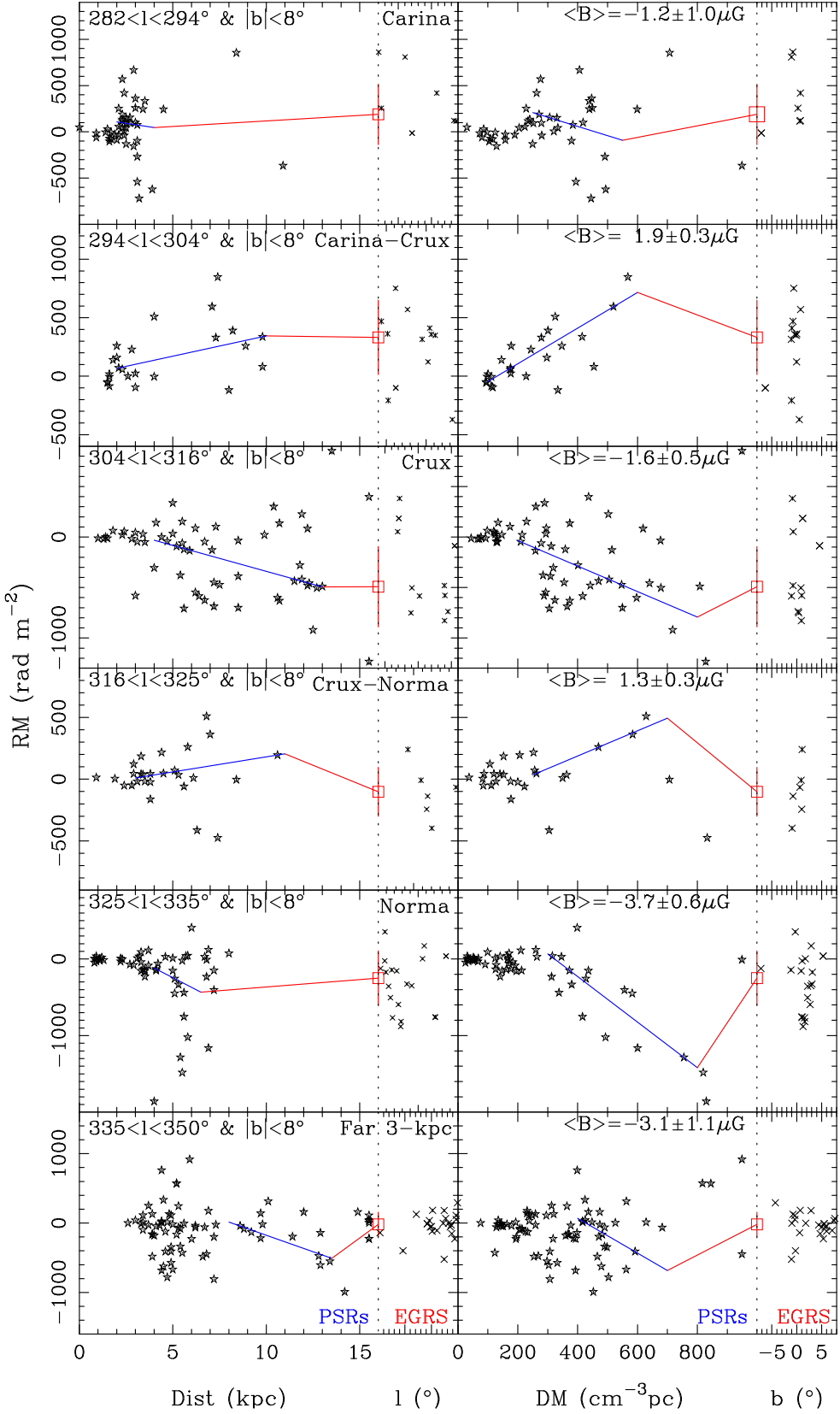}
\caption{RMs for pulsars versus pulsar distance and DM and RMs for
  EGRS versus Galactic longitude (left panels, scale ticks are given
  but not labelled) and Galactic latitude (right panels) for arm and
  interarm regions in the fourth Galactic quadrant. The blue lines
  represent linear fits to the pulsar RM gradients, primarily in the
  tangential regions. For the RM -- DM case, the slope of this line
  gives the mean interstellar line-of-sight magnetic field in the
  region with positive slopes corresponding to clockwise fields. The
  red square and error bar represent the median and rms deviation from
  the median of the RMs of EGRS in the region. Uncertainties in the
  measured RMs are plotted but are generally smaller than the plotted
  symbol size. \label{fg:rmdmq4}}
\end{figure}

We discuss Quadrant 4 first since, as viewed from the Earth, the
Galactic spiral arm and interarm regions are more clearly separated
than they are in Quadrant 1. Figure~\ref{fg:rmdmq4} shows the Quadrant
4 pulsar RMs as functions of distance and DM and EGRS RMs as a
function of Galactic latitude or longitude for the longitude ranges
given in Table~\ref{tb:tgt}.

From the top subpanels down, Figure~\ref{fg:rmdmq4} shows alternating
interarm and arm regions (cf. Table~\ref{tb:tgt}). It is striking that
the field directions alternate in the tangential regions, that is, the
RM -- DM slope is generally negative in arm regions, corresponding to
counter-clockwise field directions, but positive in interarm regions,
corresponding to clockwise field directions.

Despite the fact that there are few known pulsars beyond the
tangential zones, comparison of RMs for distant pulsars with EGRS RMs
(Figure~\ref{fg:rmdmq4}) clearly shows that for all of these longitude
zones there are field reversals beyond the tangential regions. For
example, at the far end of the Norma tangential region (325\degr $<l<$
335\degr) pulsar RMs are very negative (as much as $\sim
-1500$~rad~m$^{-2}$) whereas RMs of EGRS in this direction have a much
smaller median RM of about $-200$~rad~m$^{-2}$. This implies at least
one field reversal along this line of sight.

Similar distant reversals are seen for most of the other arm and
interarm regions. It is not possible to say exactly where these
reversals occur because of large uncertainties in the pulsar distances,
although they definitely occur beyond the fitted regions. Since both
the magnetic field strength and the electron density have a general
decline with increasing Galactocentric radius and, at least in
Quadrants 1 \& 4, the field makes a greater angle with the line of
sight beyond the tangential point, it is reasonable to assume that the
most significant zone affecting the gradient in the RMs of Galactic --
EGRS is centered on the extension of the next outer arm/interarm
region. For example, the Norma -- EGRS reversed field probably results
from the extension of the clockwise fields found in the tangential
zone of the Crux -- Norma interarm region (316\degr
$<l<$325\degr). The Crux -- Norma interarm zone itself appears to show
a reversal between the end of the tangential region and the edge of
the Galaxy, probably due to counter-clockwise fields in the distant
Crux arm. Similar considerations apply to the 3-kpc, Crux, Carina --
Crux and Carina zones although the evidence is generally somewhat
weaker compared to the Norma -- Crux zones. Table~\ref{tb:tgt} lists
the derived field strengths and directions, quantifying these field
reversals.

\begin{figure}
\includegraphics[angle=270,width=160mm]{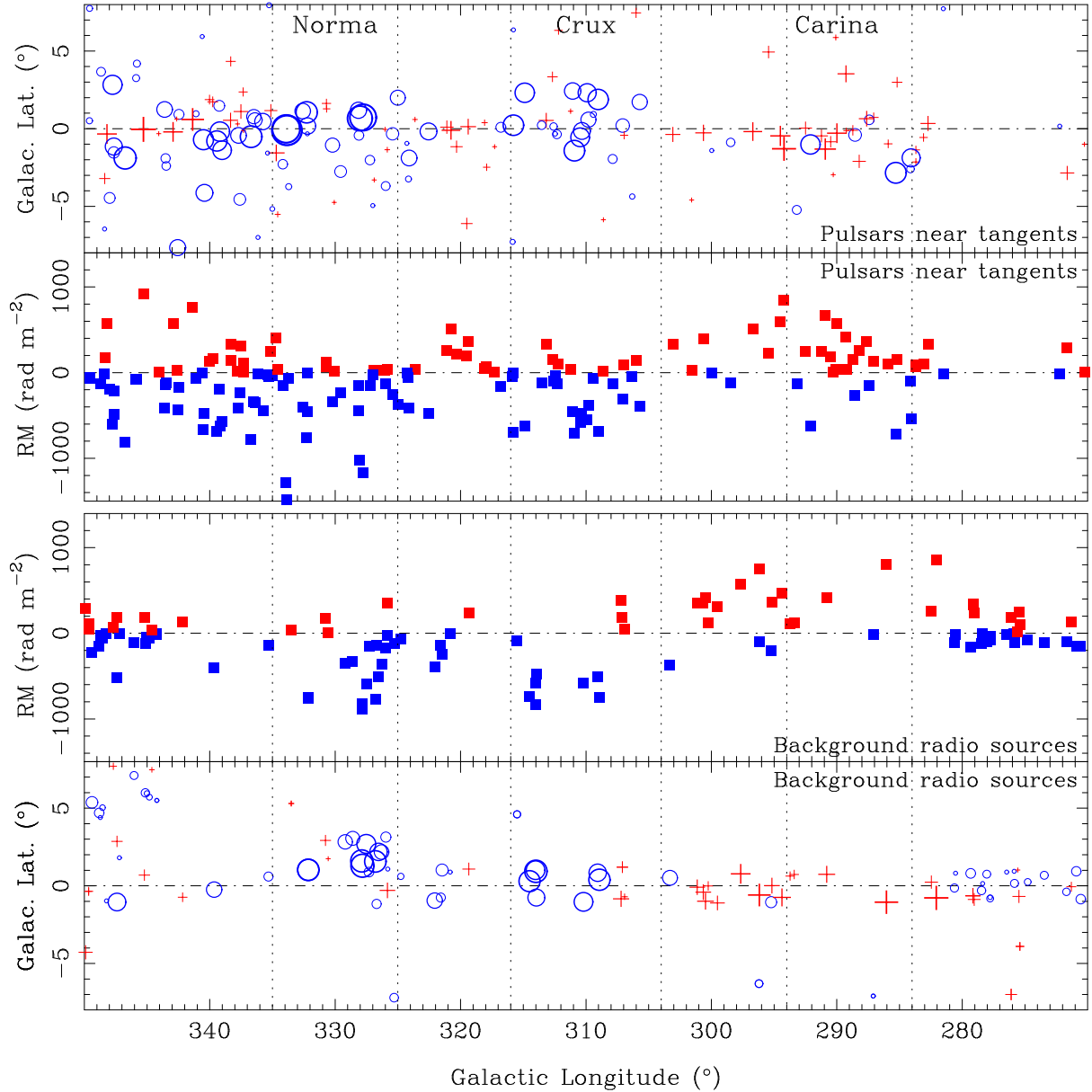}
\caption{RMs for pulsars in the tangential zone ($\pm50\%$ from the
  tangent trajectory, considering the large uncertainty of pulsar
  distances) and RMs for EGRS versus Galactic longitude in the fourth
  Galactic quadrant. The top and bottom panels show their distribution
  in Galactic latitude of the pulsars and the EGRS respectively with
  the symbol size proportional to RM$^{1/2}$.  \label{fg:RMsQ4}}
\end{figure}

These reversals are further illustrated by Figure~\ref{fg:RMsQ4} which
shows the RM distributions for pulsars in the tangential zones and for
the EGRS. The pulsar RMs (second panel) show a clear alternating
structure between arm and interarm regions at least up to the Carina
region ($l<294\degr$) with arm regions (Crux, $l\sim 310\degr$ and
Norma, $l\sim 330\degr$) predominently negative (corresponding to
counter-clockwise fields) and interarm regions predominently positive
(clockwise fields). The RMs of EGRS do not show these alternative
reversals as clearly as pulsar RMs, especially in the interarm
regions. 

\begin{figure}
\includegraphics[width=140mm]{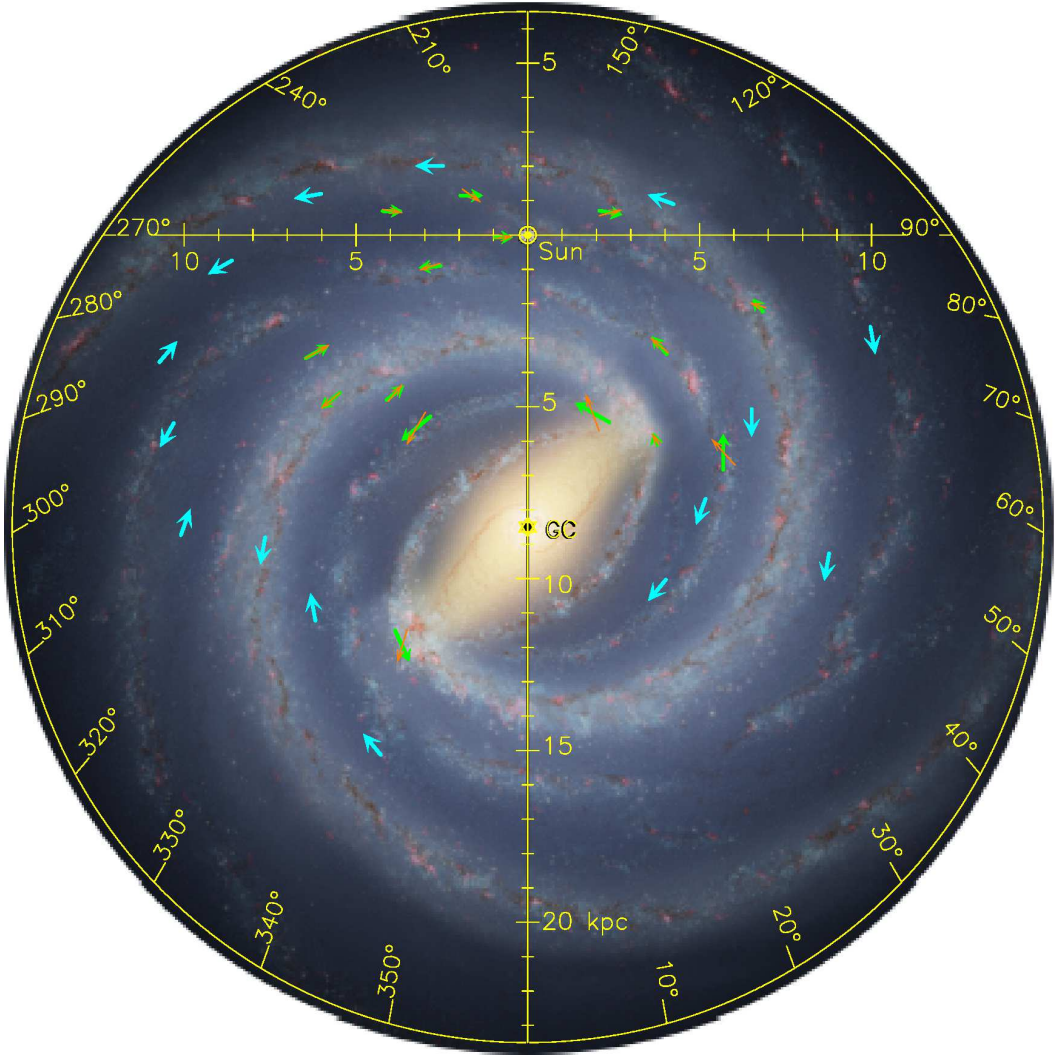}
\caption{Large-scale magnetic field directions in the Galactic disk,
  derived from the fitted gradients of pulsar RMs with DM/distance and
  the comparison of RMs of EGRS with RMs of distant pulsars. The
  orange arrows give the derived line-of-sight magnetic field
  components and the green arrows show the inferred spiral field. For
  both of these, the arrow length is proportional to the square root
  of the field strength. The blue arrows, which are all of the same
  length, give the inferred spiral field direction derived from the
  comparison of RMs for EGRS with RMs for distant pulsars and are
  placed at the distance of the central axis of the next outer spiral
  feature (arm or interarm). The background image shows an artist's
  impression of the structure of our Galaxy from a NASA/JPL image
  (Credit: R. Hurt) modified according to the spiral structure of
  \citet{hh14}.
\label{fg:disk}}
\end{figure}

The derived field strengths and directions are illustrated in
Figure~\ref{fg:disk} where the arrows are placed at the approximate
mean distance for the relevant RM -- DM fit for the pulsars and near
the next spiral feature for the pulsar -- EGRS fields as listed in
Table~\ref{tb:tgt}. As discussed above, there is substantial distance
uncertainty in both cases, but in general, the derived field
directions are consistent with reversals between the arm and interarm
regions. Arrow locations are listed in the final two columns of
Table~\ref{tb:tgt}.

\subsection{First Galactic Quadrant}\label{sec:q1}

\begin{figure}
\includegraphics[angle=270,width=140mm]{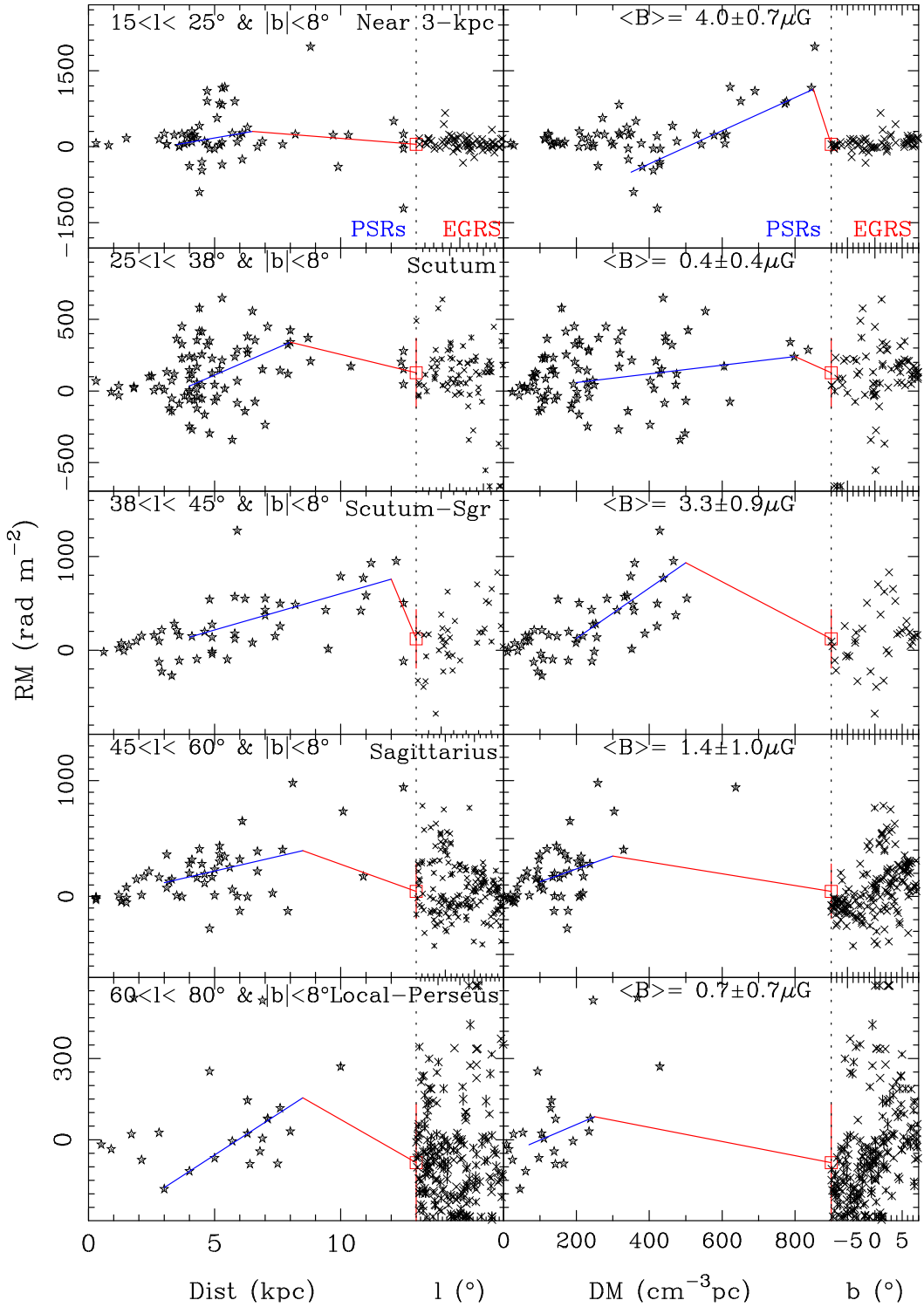}
\caption{Same as Fig.~\ref{fg:rmdmq4} but for RMs for pulsars and EGRS
  in the first Galactic quadrant. Positive slopes in the RM -- DM
  plots indicate counter-clockwise fields. \label{fg:rmdmq1}}
\end{figure}

RMs in Quadrant 1 are generally positive (see Figure~\ref{fg:rm_disk})
and there is a much less clear delineation between the arm and
interarm regions compared to Quadrant 4. Many authors have described
the positive and increasing RMs in the Sagittarius arm, implying a
counter-clockwise field in this arm
\citep[e.g.,][]{ls89,rl94,hq94,id99,wck+04}.  Figure~\ref{fg:rmdmq1}
shows positive RMs increasing with distance and DM in Sagittarius
tangential region (45\degr $<l<$ 60\degr).  The Sagittarius arm
becomes the Carina arm in Quadrant 4, supporting the idea that Carina
fields conform to the counter-clockwise pattern.

However, fields in the nominal Scutum -- Sagittarius interarm region
(38\degr $<l<$ 45\degr) show an even clearer positive and increasing
pattern, implying counter-clockwise fields in this region also. A
possible explanation for this is that these fields originate in more
distant parts (up to 12 kpc, including both the Sagittarius arm and
the Perseus arm). However, the RMs of EGRS are much smaller on average,
indicating the field reversals beyond these spiral arms.  Positive and
increasing RMs are also seen in the Near 3-kpc region (15\degr $<l<$
25\degr) which could result from the inner part of the Norma arm.

Beyond the interarm region with clockwise magnetic fields, the RM
change to positive for distant pulsars ($D>5$~kpc) in the longitude
range of 60\degr $<l<$ 80\degr in Figure~\ref{fg:rmdmq1} shows some
evidence for counter-clockwise fields in the Perseus arm. The small
RMs of EGRS in this direction indicate another field reversal beyond
the Perseus arm. The counter-clockwise field in the Perseus arm is
echoed by the RM difference of pulsars in the outer Galaxy nearer than
or within the arm and the RMs of EGRS, as we will see below.

As in Quadrant 4, there is evidence for reversals beyond the
tangential regions based on the mean RMs of EGRS, at least for the
Scutum -- Sagittarius interarm region and the Near 3-kpc region. As
discussed above, we do not know exactly where these reversals occur
because of the large DM/distance ranges for RM changes, but it is
reasonable to assume that they occur in the next arm/interarm
region. As for Quadrant 4, these field strengths and directions are
listed in Table~\ref{tb:tgt} and illustrated in Figure~\ref{fg:disk}.

\begin{figure}
\includegraphics[angle=270,width=140mm]{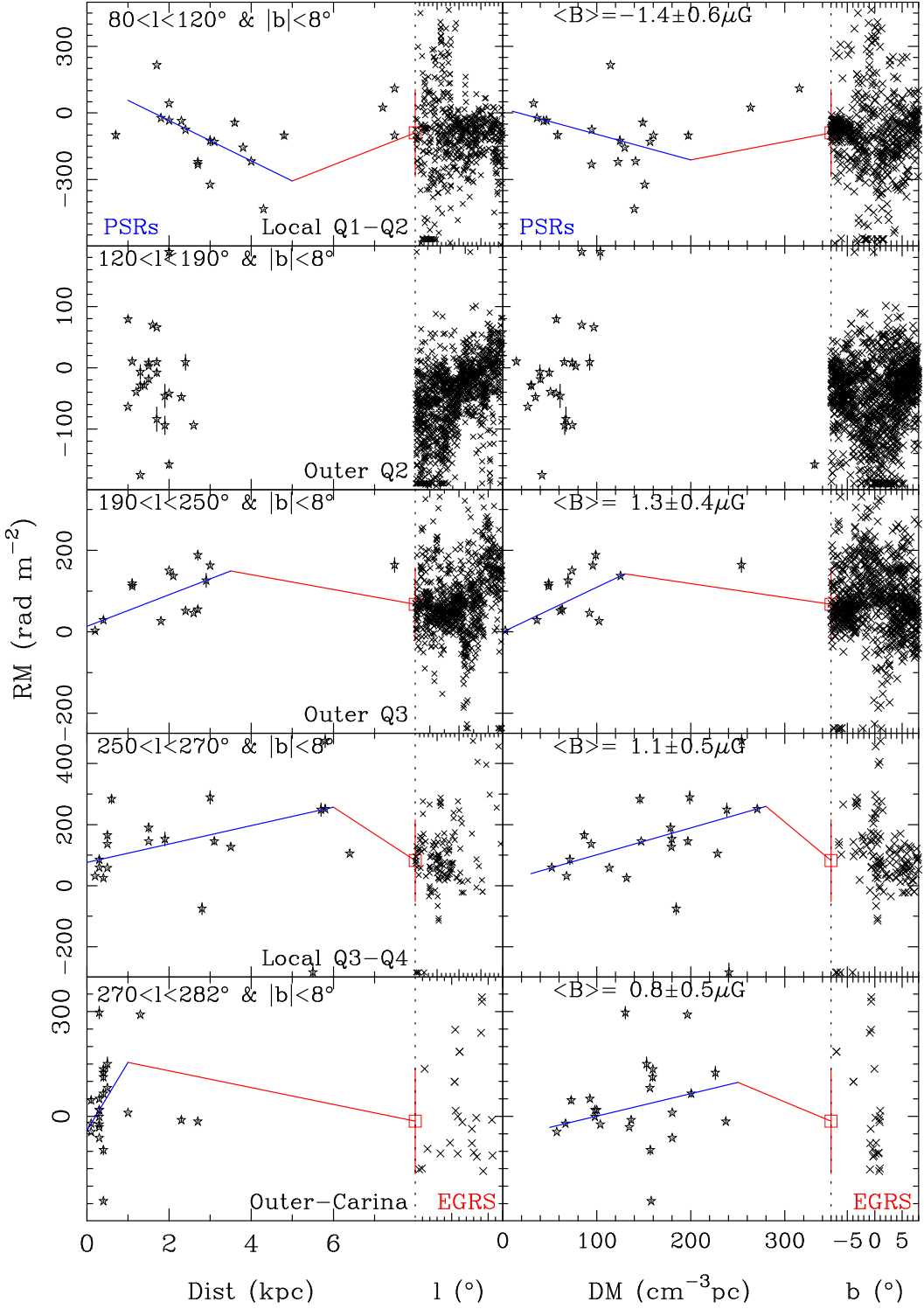}
\caption{Same as Fig.~\ref{fg:rmdmq4} but for RMs for pulsars and EGRS
  in the outer Galactic zone. For this figure, the blue lines
  represent linear fits to the change in RM for most of the known
  pulsars in each zone. Postive slopes in the RM -- DM plots indicate
  counter-clockwise fields for $l<180\degr$ and clockwise fields for
  $l>180\degr$.  \label{fg:rmdmq23}}
\end{figure}

\subsection{Outer Galactic Zones for the local interarm region and the Perseus arm}

It is well established that the Galactic magnetic field in the local
region is clockwise \citep{man74,tn80,ls89,hq94,rl94,wck+04}, implying
a reversal between the local interarm region and the Carina --
Sagittarius arm. This local clockwise field is confirmed by the
decreasing RMs of pulsars with distance and DM in the subplot in
Figure~\ref{fg:rmdmq23} for the region of $80\degr<l<120\degr$ and the
increasing RMs in the subplots for the regions of
$190\degr<l<250\degr$ and $250\degr<l<270\degr$, that is, Local
Q1--Q2, Local Q3--Q4 and also Outer-Q3, where pulsars are nearer than or
just within the Perseus arm (see Figure~\ref{fg:rm_disk}). We therefore conclude
that the clockwise fields are dominant in the Carina -- Perseus interarm
zone, including the Local Arm region.

Comparison of RMs of pulsars and EGRS gives some evidence for
counter-clockwise fields probably associated with the Perseus arm.
First of all, as seen in Figure~\ref{fg:rmdmq23}, in the region of
$80\degr<l<120\degr$, pulsars show a systematic trend for RM
decreasing. If there is no field reversal in the Perseus arm or
outside, the RMs of EGRS are expected to be more negative. However,
the data show that this is not the case. More positive RMs are
observed for not only three distant pulsars ($D>6$~kpc) but also EGRS
on average. A field reversal is also indicated by comparing
the otherwise unexpected smaller RMs of EGRS in the outer regions of
$190\degr<l<250\degr$ and $250\degr<l<270\degr$ with the increasing
RMs of pulsars.

In the longitude region of $270\degr<l<282\degr$, random but
predominantly positive RMs are observed for the local pulsars within
1~kpc, but RMs of EGRS are mostly negative which is probably an
indication of counterclockwise fields in the Perseus arm implying a
reversal from the local interarm field. If there were no field
reversal and the Perseus-arm fields were clockwise, the RMs of distant
pulsars and EGRS would be dominated by these clockwise fields and
should be positive and increasing with distance. We do not have RMs of
more distant pulsars, but the RMs of EGRS are consistent with reversed
fields in the Perseus arm.

It is difficult to probe the large-scale structure of the magnetic
field in the anti-center region of our Galaxy (e.g. in the region of
$120\degr<l<190\degr$) using Faraday rotation since the uniform field
tends to be perpendicular to the line of sight so that irregular field
fluctuations can significantly influence the measured $B_{||}$. The
RMs of pulsars in the Outer Q2 region are therefore more or less
random.

Within the limitations imposed by uncertain distances, especially for
the pulsar -- EGRS regions, the field pattern illustrated in
Figure~\ref{fg:disk} is consistent with our main conclusion, viz.,
that Galactic disk magnetic fields are predominantly counter-clockwise
in spiral arms and clockwise in interarm regions, implying field
reversals at each arm-interarm boundary. As is discussed further in
the next section, this contrasts with the field patterns derived
solely from EGRS RMs which generally have just one major region of
clockwise field encompassing the whole Carina -- Perseus region.

\section{Modeling the Galactic disk magnetic field}\label{sec:model}

Figure~\ref{fg:disk} shows the derived field directions listed in
Table~1. In general according to the analysis above, counter-clockwise
fields exist in the spiral arms and clockwise fields in the interarm
regions. Comparison of extra-galactic RMs with distant pulsar RMs
often indicate further reversals of field direction in the outer
Galaxy. Because of the uncertain electron density, it is not possible
to obtain quantitative estimates of $B_{||}$ in the regions beyond
pulsars.

For the many applications where the strength and form of the Galactic
magnetic field is important, it is useful to construct a simple model that
can reflect the field reversals we discussed above for the Galactic disk
and be used to estimate the large-scale field at a given Galactic location. Since we
have only analysed the low-latitude RMs in this paper, our model just describes
the structure of the Galactic disk field. A full three-dimensional
model is left for future work.

Our model for the Galactic disk field assumes logarithmic spiral
fields of pitch angle $\psi = 11\degr$ \citep{hh14} with a radial and
$z$ dependence given by:
\begin{equation}\label{eq:bmodel}
  B(R,z) = B_0 \exp(-R_G/A) \exp(-|z|/H)
\end{equation}
where $R_G$ is the Galactocentric radial distance, $A$ is the disk
radial scale and $H$ is the disk scale height. The field strength
$B_0 = B_s(i)$ for $R_s(i) < R_0 < R_s(i+1)$, where $R_s(i)$ and
$B_s(i)$ are given in Table~\ref{tb:bmodel} and $R_0$ is defined by
\begin{equation}
  R_0 = R_G \exp(-\psi \tan\theta)
\end{equation}
where $\theta$ is the azimuth angle measured counterclockwise from the
$+y$ axis, which points from the Galactic center to the Sun. For
$R_G <R_s(1)$ and $R_G > $ 15~kpc, we set $B_0 = 0$.

\begin{deluxetable}{lccccccc}
\tablecaption{Radial zones for the model spiral disk field. Positive values
  of $B_s$ correspond to counter-clockwise fields and negative values
  to clockwise fields, as viewed from the north Galactic pole.\label{tb:bmodel}}
\tablehead{
  \colhead{Index $i$}&\colhead{1}&\colhead{2}&\colhead{3}&\colhead{4}
  &\colhead{5}&\colhead{6}&\colhead{7}} 
\startdata
$R_s(i)$ (kpc)   & 3.0 & 4.1  & 4.9 & 6.1  & 7.5 & 8.5  & 10.5  \\
$B_s(i)$ ($\mu$G)& 4.5 &$-3.0$& 6.3 &$-4.7$& 3.3 &$-8.7$& --   \\
\enddata
\end{deluxetable}

\begin{figure}
\includegraphics[angle=270,width=180mm]{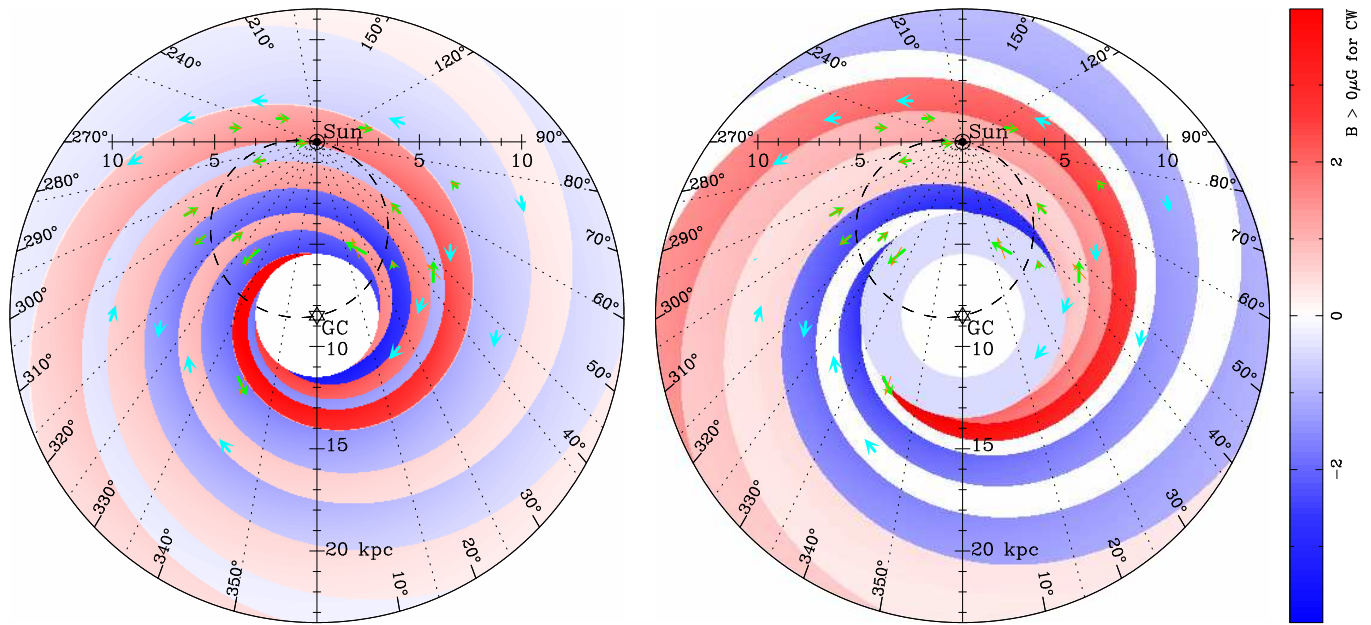}
\caption{Left panel: Model for large-scale magnetic fields in the
  Galactic disk. Blue shading represents counter-clockwise fields and
  pink shading represents clockwise fields according to the bar on the
  right side of the figure. The arrows represent the inferred spiral
  field directions as shown in Figure~\ref{fg:disk}. Right panel: The model of
  JF12 for large-scale magnetic fields in the Galactic disk
  based on RMs of EGRS and the Galactic synchrotron emission. The
  arrows are the same as in the left panel. \label{fg:models}}
\end{figure}

This field structure matches most of the field reversals we observe,
and the values of $B_s(i)$ are generally consistent with the $\langle
B\rangle$ values obtained from the gradients of the RM -- DM fits in
Figures~\ref{fg:rmdmq4} -- \ref{fg:rmdmq23}, taking into account the
location of the tangential point in Galactocentric radius and
azimuth. The scale height of the disk field, $H$, was taken to be
0.4~kpc \citep[cf.,][]{jf12} and the radial scale, $A$, was taken to
be 5.0~kpc. The model field is illustrated in the left panel of
Figure~\ref{fg:models}. The right panel of Figure~\ref{fg:models}
shows the Galactic disk field model of \citet{jf12} (JF12) which is primarily
based on RMs of EGRS.

The new model, based on a combination of pulsar RMs and EGRS RMs, has
between six and eight field reversals along a radial line from the
Galactic Center, depending on the longitude. In contrast, the
JF12 model of has clockwise disk fields across the whole
Sagittarius -- Carina region and no clockwise field in the Crux -- Norma
interarm region (tangential at $l=320\degr$). As a consequence of
this, there are only one or two reversals along radial lines from
the Galactic Center. The clockwise field in the Crux -- Norma tangential
direction is clearly indicated by the positve pulsar RMs in this zone
as shown in Figure~\ref{fg:RMsQ4}. Similarly, the pulsar RMs shown in
Figure~\ref{fg:rmdmq1} clearly show the presence of counter-clockwise
fields in the Sagittarius tangential region as discussed above in
\S\ref{sec:q1}. 

\begin{figure}
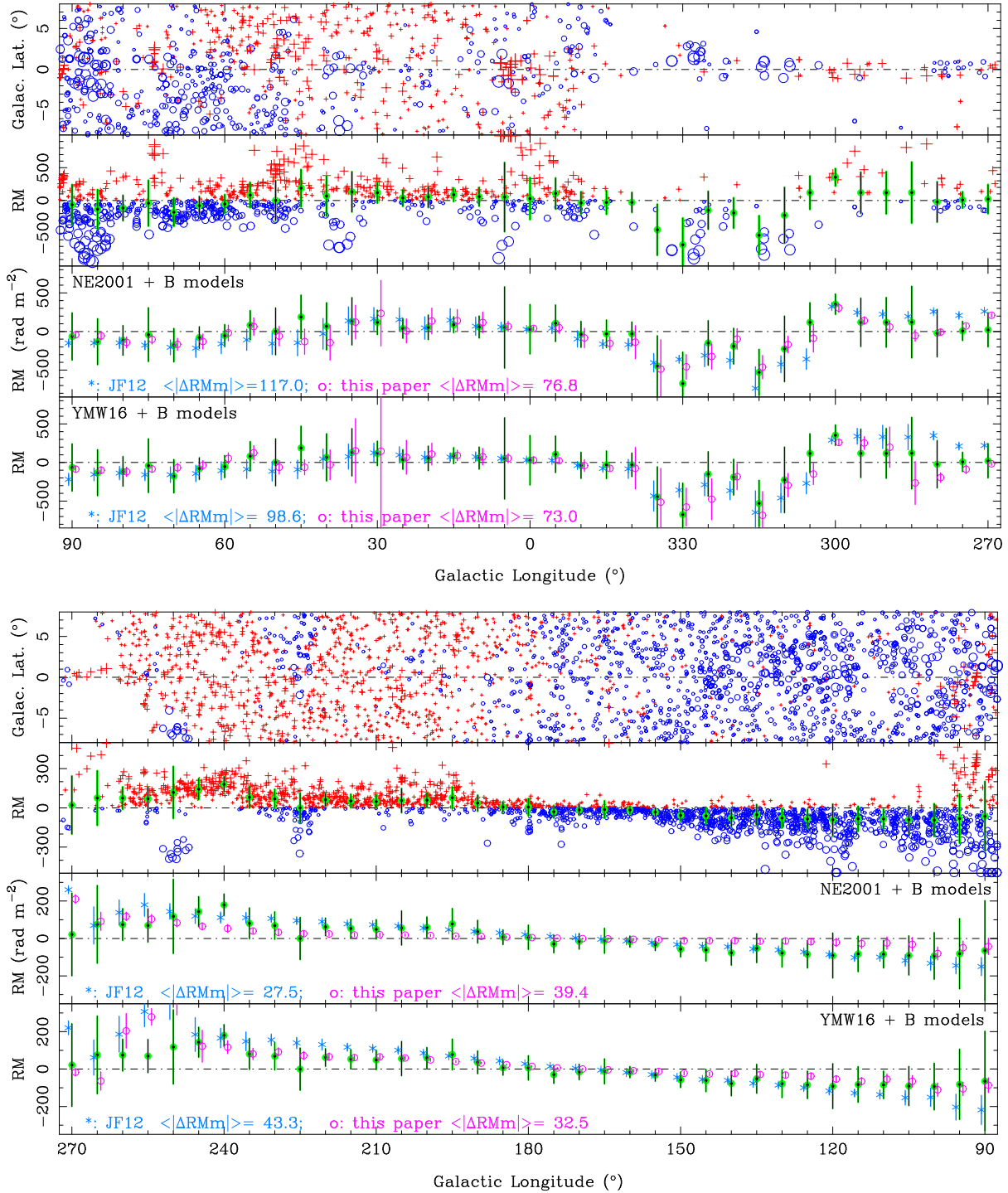

\begin{tabular}{c}
\mbox{\includegraphics[angle=270,width=160mm]{fig8a.ps}} \\
\mbox{\includegraphics[angle=270,width=160mm]{fig8b.ps}} 
\end{tabular}
\caption{Distribution of RMs of EGRS with Galactic longitude, within
  $l=0\degr \pm 90\degr$ in the upper panel and $l=180\degr \pm
  90\degr$ in the lower panel. For each panel, the top sub-panel shows
  the distribution of EGRS RMs in $b$, with blue circles representing
  RMs $<0$ and red + signs representing RMs $>0$, with the symbol size
  proportional to $|RM|^{1/2}$. The second sub-panel gives the RM
  magnitudes and their median values for every $5\degr$ of Galactic
  longitude, with error bars representing the standard deviation from
  the median in the $5\degr$ interval centered on the point. Where
  there were less than five RMs in the interval (primarily in Quadrant
  4), the interval was extended (by steps of $2\degr$) to encompass
  more than five points to give a realistic median and deviation. In
  the lower two sub-panels, the green points with error bars are as in
  the second panel, the pink circles and bars are the median values of
  the RMs computed for each source using the model for the disk field
  presented in this paper and the blue asterisks and bars are those
  computed using the JF12 model, for the NE2001 electron
  density model \citep{cl02} and the YMW16 model \citep{ymw17},
  respectively. The model points and bars are slightly offset for
  clarity.  Mean absolute differences between the median observed RMs
  and median predicted RMs weighted by the rms deviation of the
  observed RMs, $\langle |\Delta RM_m |\rangle$, is given for each
  field model and each electron density model in the lower
  sub-panels. \label{fg:egrmgl}}
\end{figure}

To compare how well these two models ``predict'' the EGRS RMs, we
compute the RM of each EGRS source in Figure~\ref{fg:rm_disk} using
both the NE2001 electron density model \citep{cl02}, as used by JF12,
and the more recent YMW16 model \citep{ymw17}. In
Figure~\ref{fg:egrmgl} we show the distribution of observed EGRS RMs
in Galactic latitude and longitude, along with the RMs computed using
our model for the disk field, which is based on both pulsar and EGRS
RMs, and from the JF12 model which is based on just the EGRS RMs.  The
RM median values for every $5\degr$ of Galactic longitude are compared
in the lower two sub-panels. This figure demonstrates that, despite its
different structure and different basis, our model predicts RMs of
EGRS more accurately than the JF12 model which is based primarily on
them, for example, in the region $270\degr \ge l \ge 280\degr$. With
the YMW16 electron density model \citep{ymw17}, the weighted mean
difference between the median observed RMs and median predicted RMs is
smaller for our model in all quadrants. Even with the NE2001 model,
our model is significantly better than predictions based on the JF12
model for the first and fourth quadrants (i.e., top panel of
Figure~\ref{fg:egrmgl}).

However, the converse is not true. For example, as shown in the right
panel of Figure~\ref{fg:models}, the EGRS-based model does not
correctly model the clockwise fields in the Crux-Norma interarm region
or the counter-clockwise fields in the Sagittarius-Carina arm. With
their distribution through the Galactic disk at approximately known
distances enabling tomographic mapping, pulsars are able to reveal
reversals in the large-scale disk field that are concealed in the RMs
of EGRS since they integrate across the entire disk. Despite the large
number of EGRS RMs in the outer regions of our Galaxy (i.e. within
$l=180\degr \pm 90\degr$), the field structure is not well constrained
without comparison with the RMs of foreground pulsars. Even though the
number of pulsar RMs is relatively small compared to that of EGRS RMs,
the pulsar RMs are a very powerful tool in investigations of the
large-scale Galactic magnetic field.

\section{Conclusions}\label{sec:concl}

We have measured rotation measures for 477 pulsars of which 441 are
either new or improved over previous measurements. By analyzing the
distribution of pulsar RMs and comparing RMs for pulsars and
extra-galactic radio sources (EGRS) lying within $8\degr$ of the
Galactic Plane, we show that the large-scale disk field in the inner
Galaxy probably has a bisymmetric form with reversals between spiral
arm and interarm regions. Compared to the analysis in \citet{hml+06},
we have a larger sample of pulsar RMs and have combined pulsar
and EGRS data to show the reversals in the Galactic disk large-scale
field more clearly. Most of these reversals are not apparent in EGRS
RM measurements since these average over the whole path inside our
Galaxy. Furthermore, pulsar RM and DM data can give direct
measurements of the mean magnetic field strength in selected regions
of the Galaxy, for example, zones around tangential points.

Based on these results, we present a quantitative model for the
large-scale magnetic field in the Galactic disk, which not only models
the spiral magnetic field reversals between arm and interarm regions,
but also can reproduce the RM distribution of EGRS better than a
recent model based on EGRS data alone. 

In the future, more pulsar RMs and improved pulsar distances will
become available, allowing the large-scale structure of the Galactic
magnetic fields to be better constrained.  More RMs of EGRS, particularly
in the longitude zone from $260\degr$ to $350\degr$ can help to
determine the magnetic field structure beyond the
pulsars. Observations of higher-latitude RMs for both pulsars and EGRS
may in future allow construction of a complete three-dimensional model
of the large-scale Galactic magnetic field, which is a proposed
project for the SKA \citep[see e.g.,][]{hvl+15}.

\section*{ACKNOWLEDGMENTS}
We sincerely thank Dr. Jun Xu for help on construction of the models
for the disk magnetic field, and Dr. LiGang Hou for help on the
background spiral images. JLH is supported by the Key Research Program
of the Chinese Academy of Sciences (Grant No. QYZDJ-SSW-SLH021) and
the National Natural Science Foundation (No. 11473034). The Parkes
radio telescope is part of the Australia Telescope which is funded by
the Commonwealth Government for operation as a National Facility
managed by the Commonwealth Scientific and Industrial Research
Organisation. The National Radio Astronomy Observatory is a facility
of the National Science Foundation operated under cooperative
agreement by Associated Universities, Inc.


\end{document}

%% file: tab1.tex
\begin{deluxetable}{lrrrrrrrcc}                                                                      
\tabletypesize{\footnotesize}                                                                        
\tablecolumns{10}                                    
\tablewidth{0pt}             
\tablecaption{Observed rotation measures for 477 pulsars\label{tb:psrrm}}    
\tablehead{          
\colhead{PSR Name} & \colhead{Period} & \colhead{DM} &       
\colhead{Gal. $l$} & \colhead{Gal. $b$} & \colhead{Dist.} &          
\colhead{RM} & \colhead{$\sigma_{\rm RM}$} &         
\colhead{Telescope} & \colhead{Obs. Date} \\         
 & \colhead{(s)} & \colhead{(cm$^{-3}$)} &           
\colhead{(\degr)} & \colhead{(\degr)} & \colhead{(kpc)} &            
\colhead{(rad m$^{-2}$)} & \colhead{(rad m$^{-2}$)} & &      
}            
\startdata           
J0014$+$4746  &  1.2407 &   30.85 &  116.50 &$ -14.63$ &    1.78 &$  -15.3 $ &  0.7 & GBT & 071119 \\ 
J0030$+$0451  &  0.0049 &    4.33 &  113.14 &$ -57.61$ &    0.36 &$   16.8 $ & 15.9 & PKS & 080112 \\ 
J0034$-$0534  &  0.0019 &   13.76 &  111.49 &$ -68.07$ &    1.35 &$  -38.1 $ & 17.5 & PKS & 080113 \\ 
J0034$-$0721  &  0.9430 &   11.38 &  110.42 &$ -69.81$ &    1.03 &$    3.9 $ & 10.4 & PKS & 080215 \\ 
J0055$+$5117  &  2.1152 &   44.12 &  123.62 &$ -11.58$ &    1.94 &$  -66.6 $ &  1.5 & GBT & 071118 \\ 
J0113$-$7220  &  0.3259 &  125.49 &  300.62 &$ -44.69$ &   59.70 &$   87.0 $ & 28.7 & PKS & 080111 \\ 
J0117$+$5914  &  0.1014 &   49.42 &  126.28 &$  -3.46$ &    1.77 &$   -8.1 $ &  6.7 & GBT & 071118 \\ 
\enddata 
\tablecomments{Table \ref{tb:psrrm} is published in its entirety in the machine-readable format.
      A portion is shown here for guidance regarding its form and content.}
\end{deluxetable}

%% file: tab2.tex
\begin{deluxetable}{lrrlrr}
\tabletypesize{\footnotesize}
\tablecolumns{6}
\tablewidth{0pt}
\tablecaption{Weighted mean rotation measures\label{tb:meanrm}}
\tablehead{
\colhead{PSR Name} & \colhead{RM} & \colhead{$\sigma_{\rm RM}$} &
\colhead{PSR Name} & \colhead{RM} & \colhead{$\sigma_{\rm RM}$} 
}
\startdata
J0437$-$4715 &$    0.4  $& 0.2  & J1730$-$2304 &$   -4.9  $& 1.8  \\
J0656$-$2228 &$   83.0  $& 5.4  & J1759$-$2302 &$ 1574.7  $& 13.0 \\
J0815$+$0939 &$   53.1  $& 5.0  & J1812$-$2102 &$  322.6  $& 4.2  \\
J0842$-$4851 &$  145.1  $& 11.3 & J1816$-$1729 &$   82.9  $& 4.1  \\
J0900$-$3144 &$   82.0  $& 1.3  & J1843$-$0355 &$  239.7  $& 9.8  \\
J0902$-$6325 &$ -59.2   $& 2.1  & J1852$+$0305 &$  263.9  $& 14.6 \\
J0942$-$5552 &$ -63.2   $& 1.4  & J1915$+$0738 &$   -6.8  $&  1.5 \\
J0952$-$3839 &$  331.7  $& 9.5  & J1927$+$1856 &$   74.4  $&  5.8 \\
J1032$-$5911 &$  100.0  $& 6.7  & J1932$+$1059 &$   -7.5  $&  0.4 \\
J1626$-$4537 &$  111.9  $& 11.2 & J2053$-$7200 &$   15.2  $&  2.0 \\
J1707$-$4053 &$ -186.5  $& 2.5  \\
\enddata
\end{deluxetable}

%% file: tab3.tex
\begin{deluxetable}{llrrrrlrrlrrl}
\tabletypesize{\footnotesize}
\tablecolumns{13}
\tablewidth{0pt}
\tablecaption{Comparison of observed rotation measures with previously published values\label{tb:pubrm}}
\tablehead{
\multicolumn{2}{c}{PSR Name} &
\colhead{RM} & \colhead{$\sigma_{\rm RM}$} &
\colhead{RM$_1$} & \colhead{$\sigma_{\rm RM_1}$} & \colhead{Ref. 1} &
\colhead{RM$_2$} & \colhead{$\sigma_{\rm RM_2}$} & \colhead{Ref. 2} &
\colhead{RM$_3$} & \colhead{$\sigma_{\rm RM_3}$} & \colhead{Ref. 3} 
}
\startdata
J0014$+$4746 & B0011$+$47 & $   -15.3$ &   0.7 &$   -8.7 $& 1.1  &  fdr15 &$       $ &      &       &$         $&      &       \\
J0034$-$0721 & B0031$-$07 & $     3.9$ &  10.4 &$   9.89 $& 0.07 & nsk+15*&$    9.8 $&  0.2 & hl87  &$    10.0 $&  1.0 & man74 \\
J0437$-$4715 & \nodata    & $     0.4$ &   0.2 &$    1.5 $&  0.5 & nms+97 &$    0.0$ &  0.4 & ymv+11&$         $&      &       \\
J0448$-$2749 & \nodata    & $    -0.7$ &   5.4 &$   24.0 $& 17.0 & hml+06 &$       $ &      &       &$         $&      &       \\
J0452$-$1759 & B0450$-$18 & $    12.7$ &   0.9 &$   11.1 $&  0.3 & jkk+07*&$   13.8$ &  0.7 &  hl87 &$    15.0 $&  2.0 & man74 \\
J0536$-$7543 & B0538$-$75 & $    23.8$ &   0.9 &$   25.2 $&  1.0 & njkk08 &$   28.0$ &  2.0 & hmq99 &$    21.4 $&  0.5 & qmlg95\\
J0630$-$2834 & B0628$-$28 & $    45.4$ &   0.7 &$   46.5 $&  0.1 & jhv+05*&$   46.6$ &  1.3 & hml+06&$    45.7 $&  0.5 & vdhm97\\
J0656$-$2228 & \nodata    & $    83.0$ &   5.4 &$   38.0 $& 12.0 & njkk08 &$       $ &      &       &$         $&      &       \\
J0738$-$4042 & B0736$-$40 & $     9.3$ &   0.9 &$   12.1 $&  0.6 & njkk08*&$   14.5$ &  0.7 & vdhm97&$   12.5  $&  0.6 & jkk+07\\
J0831$-$4406 & \nodata    & $   531.4$ &  22.9 &$  509.0 $& 20.0 & hml+06 &$       $ &      &       &$         $&      &       \\
J0835$-$4510 & B0833$-$45 & $    29.9$ &   0.6 &$  31.38 $& 0.01 & jhv+05*&$   36.6$ &  0.1 & man74 &$   38.2  $&  0.1 & hmm+77\\
J0838$-$2621 & \nodata    & $    96.3$ &  12.4 &$   86.0 $& 13.0 & njkk08 &$       $ &      &       &$         $&      &       \\
J0843$-$5022 & \nodata    & $   189.5$ &  15.9 &$  155.0 $& 23.0 & njkk08 &$       $ &      &       &$         $&      &       \\
J0846$-$3533 & B0844$-$35 & $   136.5$ &   3.5 &$  144.0 $&  8.0 & hl87   &$  159.0$ &  9.0 & qmlg95&$         $&      &       \\
J0942$-$5552 & B0940$-$55 & $   -62.8$ &   1.8 &$  -61.9 $&  0.2 & tml93* &$       $ &      &       &$         $&      &       \\
J0953$+$0755 & B0950$+$08 & $     6.1$ &   1.6 &$  -0.66 $& 0.04 & jhv+05*&$    2.0$ &  2.0 & hl87  &$     1.8 $&  0.5 & man74 \\
J1012$+$5307 & \nodata    & $     1.0$ &   1.4 &$   2.98 $& 0.06 & nsk+15*&$       $ &      &       &$         $&      &       \\
J1017$-$5621 & B1015$-$56 & $   332.8$ &   3.6 &$  365.0 $&  7.0 & njkk08 &$       $ &      &       &$         $&      &       \\
J1022$+$1001 & \nodata    & $     8.6$ &   9.5 &$   1.39 $& 0.05 & nsk+15*&$  -0.6 $&  0.5 & ymv+11 &$         $&      &       \\
J1024$-$0719 & \nodata    & $    -5.8$ &   3.5 &$   -8.2 $&  0.8 & ymv+11*&$       $ &      &       &$         $&      &       \\
J1045$-$4509 & \nodata    & $    90.5$ &   3.5 &$   92.0 $&  1.0 & ymv+11*&$   82.0$ & 18.0 & mh04  &$         $&      &       \\
J1047$-$3032 & \nodata    & $   -26.2$ &   6.2 &$  -36.0 $& 23.0 & njkk08 &$       $ &      &       &$         $&      &       \\
J1052$-$5954 & \nodata    & $  -269.7$ &  11.7 &$ -280.0 $& 24.0 & wj08   &$       $ &      &       &$         $&      &       \\
J1054$-$5943 & \nodata    & $   150.6$ &  30.1 &$   46.0 $& 34.0 & hml+06 &$       $ &      &       &$         $&      &       \\
J1115$-$6052 & \nodata    & $   251.6$ &   4.6 &$  257.0 $& 18.0 & wj08   &$       $ &      &       &$         $&      &       \\
J1156$-$5707 & \nodata    & $   228.0$ &   6.3 &$  238.0 $& 19.0 & wj08   &$       $ &      &       &$         $&      &       \\
J1237$-$6725 & \nodata    & $   72.5 $ &  21.8 &$   24.0 $& 14.0 & tjb+13*&$       $ &      &       &$         $&      &       \\
J1240$-$4124 & B1237$-$41 & $    17.3$ &  10.4 &$   15.0 $& 13.0 & njkk08 &$       $ &      &       &$         $&      &       \\
J1300$+1$240 & B1257+12   & $     2.3$ &   3.5 &$   7.91 $& 0.06 & nsk+15*&$       $ &      &       &$         $&      &       \\
J1312$-$6400 & \nodata    & $   -12.9$ &   9.1 &$   40.0 $& 30.0 & hml+06 &$       $ &      &       &$         $&      &       \\
J1320$-$3512 & \nodata    & $    -4.7$ &   3.6 &$   -7.8 $&  2.0 & njkk08*&$       $ &      &       &$         $&      &       \\
J1321$+$8323 & B1322+83   & $   -27.5$ &   1.4 &$  -23.1 $&  1.1 &  fdr15*&$       $ &      &       &$         $&      &       \\
J1340$-$6456 & B1336$-$64 & $    -2.1$ &  20.8 &$  -37.0 $& 23.0 & njkk08 &$       $ &      &       &$         $&      &       \\
J1352$-$6803 & \nodata    & $    20.4$ &   2.4 &$   30.0 $&  7.0 & njkk08 &$       $ &      &       &$         $&      &       \\
J1403$-$7646 & \nodata    & $    63.4$ &  17.3 &$   94.0 $& 16.0 & njkk08*&$       $ &      &       &$         $&      &       \\
J1514$-$4834 & B1510$-$48 & $    13.8$ &  18.2 &$   18.0 $& 14.0 & njkk08*&$       $ &      &       &$         $&      &       \\
J1524$-$5625 & \nodata    & $   185.8$ &   3.3 &$  180.0 $& 20.0 & wj08   &$       $ &      &       &$         $&      &       \\
J1524$-$5706 & \nodata    & $  -475.5$ &   5.0 &$ -470.0 $& 20.0 & wj08   &$       $ &      &       &$         $&      &       \\
J1534$-$5405 & B1530$-$53 & $   -86.8$ &   6.0 &$  -69.0 $& 12.0 & njkk08 &$       $ &      &       &$         $&      &       \\
J1600$-$3053 & \nodata    & $   -11.9$ &   3.7 &$  -15.5 $&  1.0 & ymv+11*&$       $ &      &       &$         $&      &       \\
J1614$-$3937 & \nodata    & $    84.0$ &  13.7 &$  133.0 $& 16.0 & njkk08 &$       $ &      &       &$         $&      &       \\
J1615$-$5537 & B1611$-$55 & $    10.2$ &  14.8 &$  -54.0 $& 16.0 & njkk08 &$       $ &      &       &$         $&      &       \\
J1623$-$4256 & B1620$-$42 & $  109.6 $ &  5.7  &$  -15.0 $&  8.0 & hml+06 &$       $ &      &       &$         $&      &       \\
J1628$-$4804 & \nodata    & $  -447.0$ &  15.5 &$ -431.0 $& 43.0 & hml+06 &$       $ &      &       &$         $&      &       \\
J1644$-$4559 & B1641$-$45 & $  -626.9$ &   0.8 &$ -617.0 $&  1.0 & hml+06 &$ -611.0$ &  2.0 & vdhm97&$         $&      &       \\
J1650$-$1654 & \nodata    & $    16.1$ &   5.6 &$    7.0 $& 14.0 & njkk08 &$       $ &      &       &$         $&      &       \\
J1651$-$4246 & B1648$-$42 & $  -167.4$ &   1.1 &$ -154.0 $&  5.0 & hml+06 &$       $ &      &       &$         $&      &       \\
J1702$-$4128 & \nodata    & $  -165.5$ &   4.7 &$ -160.0 $& 20.0 & wj08   &$       $ &      &       &$         $&      &       \\
J1705$-$3950 & \nodata    & $   -98.8$ &   1.9 &$ -106.0 $& 14.0 & wj08   &$       $ &      &       &$         $&      &       \\
J1707$-$4053 & B1703$-$40 & $  -183.7$ &   3.4 &$  168.0 $&  4.0 & njkk08 &$ -207.0$ & 25.0 & qmlg95&$         $&      &       \\
J1709$-$4429 & B1706$-$44 & $    -3.5$ &   0.8 &$   0.70 $& 0.07 & jhv+05*&$   -7.0$ &  4.0 & qmlg95&$         $&      &       \\
J1713$+$0747 & \nodata    & $    7.0 $ &   7.6 &$    8.4 $&  0.6 & ymv+11*&$       $ &      &       &$         $&      &       \\ 
J1717$-$4054 & B1713$-$40 & $  -811.4$ &   4.1 &$ -800.0 $&100.0 & khs+14 &$       $ &      &       &$         $&      &       \\ 
J1721$-$3532 & B1718$-$35 & $   148.1$ &   5.9 &$  159.0 $&  4.0 & njkk08*&$  205.0$ & 75.0 & qmlg95&$         $&      &       \\
J1730$-$2304 & \nodata    & $    -4.9$ &   1.8 &$   -7.2 $&  2.2 & ymv+11 &$       $ &      &       &$         $&      &       \\
J1737$-$3137 & \nodata    & $   449.6$ &   5.1 &$  448.0 $& 17.0 & wj08   &$       $ &      &       &$         $&      &       \\
J1737$-$3555 & B1734$-$35 & $    68.6$ &   9.8 &$   50.0 $&  4.0 & njkk08*&$       $ &      &       &$         $&      &       \\
J1744$-$1134 & \nodata    & $     6.0$ &   2.1 &$   -1.6 $&  0.7 & ymv+11*&$       $ &      &       &$         $&      &       \\
J1818$-$1422 & B1815$-$14 & $  1173.9$ &   7.3 &$ 1168.0 $& 13.0 & hml+06 &$       $ &      &       &$         $&      &       \\
J1822$-$4209 & \nodata    & $    40.7$ &  10.6 &$  -13.0 $&  9.0 & hml+06 &$       $ &      &       &$         $&      &       \\
J1828$-$1101 & \nodata    & $    59.4$ &   2.5 &$   45.0 $& 20.0 & wj08   &$       $ &      &       &$         $&      &       \\
J1835$-$0643 & B1832$-$06 & $    44.1$ &  14.6 &$   62.0 $& 38.0 & hml+06 &$       $ &      &       &$         $&      &       \\
J1835$-$1106 & \nodata    & $    42.9$ &   2.1 &$   42.0 $&  3.0 & njkk08 &$       $ &      &       &$         $&      &       \\
J1836$-$1008 & B1834$-$10 & $   826.6$ &   4.5 &$-1000.0 $& 99.0 & hl87   &$       $ &      &       &$         $&      &       \\
J1837$-$0045 & \nodata    & $   131.6$ &   6.7 &$  130.0 $& 17.0 & njkk08 &$       $ &      &       &$         $&      &       \\
J1837$-$0604 & \nodata    & $   320.8$ &   4.2 &$  450.0 $& 25.0 & wj08   &$       $ &      &       &$         $&      &       \\
J1837$-$1837 & \nodata    & $   137.2$ &   9.0 &$  138.0 $&  8.0 & njkk08*&$       $ &      &       &$         $&      &       \\
J1841$-$0345 & \nodata    & $   450.5$ &   2.6 &$  447.0 $& 15.0 & wj08   &$       $ &      &       &$         $&      &       \\
J1845$-$0743 & \nodata    & $   448.4$ &   1.8 &$  440.0 $& 12.0 & wj08   &$       $ &      &       &$         $&      &       \\
J1853$-$0004 & \nodata    & $   648.7$ &   4.7 &$  647.0 $& 16.0 & wj08   &$       $ &      &       &$         $&      &       \\
J1900$-$2600 & B1857$-$26 & $    -9.3$ &   0.2 &$   -2.3 $&  0.8 & jhv+05 &$   -7.3$ &  0.8 & hl87  &$         $&      &       \\
J1900$-$7951 & B1851$-$79 & $    18.6$ &   5.4 &$   43.0 $& 12.0 & qmlg95 &$       $ &      &       &$         $&      &       \\
J1901$-$1740 & \nodata    & $   -20.0$ &   8.7 &$   63.0 $& 33.0 & njkk08 &$       $ &      &       &$         $&      &       \\
J1903$+$0135 & B1900$+$01 & $    68.4$ &   2.4 &$   72.3 $&  1.0 & hl87*  &$       $ &      &       &$         $&      &       \\
J1915$+$1606 & B1913$+$16 & $   364.5$ &   5.0 &$  430.0 $& 73.0 & hml+06 &$       $ &      &       &$         $&      &       \\
J1917$+$2224 & B1915$+$22 & $   168.2$ &  23.5 &$  192.0 $& 49.0 & wck+04 &$       $ &      &       &$         $&      &       \\
J1919$+$0134 & \nodata    & $    44.8$ &  18.2 &$   47.0 $&  4.0 & njkk08*&$       $ &      &       &$         $&      &       \\
J1921$+$1419 & B1919+14   & $   164.8$ &   3.1 &$  275.0 $& 60.0 & hr10   &$       $ &      &       &$         $&      &       \\
J1926$+$0431 & B1923$+$04 & $   -39.5$ &   8.1 &$    0.0 $& 11.0 & hl87   &$       $ &      &       &$         $&      &       \\
J1932$+$1059 & B1929$+$10 & $    -8.2$ &   0.8 &$  -6.87 $& 0.02 & jhv+05*&$   -7.0$ &  2.0 & hl87  &$    -8.6 $&  1.8 & man74 \\
J1932$-$3655 & \nodata    & $     3.8$ &   4.7 &$   -8.0 $&  3.0 & njkk08*&$   -6.0$ &  3.0 & hmq99 &$         $&      &       \\
J1935$+$1616 & B1933$+$16 & $    -2.3$ &   0.5 &$  -10.2 $&  0.3 & jhv+05 &$   -2.0$ &  2.0 & hl87  &$    -1.9 $&  0.4 & man74 \\
J1943$-$1237 & B1940$-$12 & $   -75.4$ &   6.2 &$  -10.0 $&  8.0 & hl87   &$       $ &      &       &$         $&      &       \\
J1949$-$2524 & B1946$-$25 & $   -51.9$ &  13.8 &$  -13.0 $&  8.0 & hmq99  &$       $ &      &       &$         $&      &       \\
J2038$-$3816 & \nodata    & $    20.5$ &   9.5 &$   38.0 $& 14.0 & njkk08 &$   68.0$ & 18.0 & hml+06&$         $&      &       \\
J2048$-$1616 & B2045$-$16 & $    -9.7$ &   1.7 &$  -10.0 $&  0.3 & jkk+07*&$   -9.0$ &  2.0 & hl87  &$   -10.8 $&  0.4 & man74 \\
J2053$-$7200 & B2048$-$72 & $    15.0$ &   2.0 &$   17.0 $&  1.0 & qmlg95*&$       $ &      &       &$         $&      &       \\
J2108$-$3429 & \nodata    & $    93.1$ &   8.9 &$   39.0 $& 12.0 & njkk08 &$   50.0$ & 20.0 & hmq99 &$         $&      &       \\
J2155$-$3118 & B2152$-$31 & $    33.8$ &  14.4 &$   21.0 $&  3.0 & hl87*  &$       $ &      &       &$         $&      &       \\
J2324$-$6054 & B2321$-$61 & $    15.6$ &   2.0 &$  -11.0 $&  8.0 & hml+06 &$   39.0$ &  6.0 & qmlg95&$         $&      &       \\
\hline
\multicolumn{13}{l}{\parbox[t]{17cm}{All RM values and their uncertainties
are in units of rad~m$^{-2}$. \\
References marked with * signify the best available RM values if not from 
the present work. \\
References: 
fdr15: \citet{fdr15};
hl87: \citet{hl87}; 
hml+06: \citet{hml+06};
hmm+77: \citet{hmm+77};
hmq99: \citet{hmq99}; 
hr10: \citet{hr10};
jhv+05: \citet{jhv+05};
jkk+07: \citet{jkk+07};
khs+14: \citet{khs+14};
man74: \citet{man74};
mh04: \citet{mh04};
njkk08: \citet{njkk08};
nms+97: \citet{nms+97};
nsk+15: \citet{nsk+15};
qmlg95: \citet{qmlg95};
tjb+13: \citet{tjb+13};
tml93: \citet{tml93}; 
vdhm97: \citet{vdhm97}; 
wck+04: \citet{wck+04};
wj08: \citet{wj08};
ymv+11: \citet{ymv+11}
}}
\enddata
\end{deluxetable}